\documentclass[twocolumn,showpacs,preprintnumbers,amsmath,amssymb,pra]{revtex4}
\usepackage{graphicx}
\usepackage{psfig}
\usepackage{graphics}
\usepackage{amsmath}
\usepackage{dcolumn}
\usepackage{amssymb}
\usepackage{bm}
\begin{document}
%data:21/01/2003

\title{Quantum Dynamics of Magnetic and Electric Dipoles and
Berry's Phase}
\author{ Claudio  Furtado }
\affiliation{ Departamento de F\'{\i}sica,\\ Universidade Federal da Para\'{\i}ba,Caixa Postal 5008, 58051-970, Jo\~ao Pessoa, PB, Brazil}
\author{ C. A. de Lima Ribeiro}
\affiliation{ Departamento de F\'{\i}sica,\\Universidade Estadual de Feira de
Santana\\44031-460, Feira de Santana, BA, Brazil}
\date{\today}
\begin{abstract}
We study the quantum dynamics of neutral particle that posseses a permanent magnetic and electric  dipole  moments in the presence of an electromagnetic field. The analysis of this dynamics demonstrates the appearance of a quantum phase that combines the Aharonov-Casher effect and the He-Mckellar-Wilkens effect. We demonstrate that this phase is a special case of the Berry's quantum phase. A series of field configurations where this phase would be found are presented.  A generalized Casella-type effect is found in one these configurations. A physical
scenario for the quantum  phase in an interferometric experiment is proposed.

\end{abstract}

\maketitle

In 1959 Aharonov and Bohm (AB)\cite{aha} demonstrated that a
quantum charge circulating a  magnetic flux tube acquire
a quantum topological phase. This effect was observed
experimentally by Chambers\cite{prl:cham,pes}. Aharonov and
Casher(AC) showed that a particle with a magnetic
moment moving in an electric field accumulates a quantum phase,
the AC phase\cite{cas}. This phenomenon has been
observed in a neutron interferometer\cite{cim} and in a
neutral atomic Ramsey interferometer\cite{san,san1}. 

He and McKellar\cite{mac}, and Wilkens\cite{wil}, independently,
have predicted the existence of a quantum phase that an electric
dipole acquires, while circulating around, and parallel to a
line of magnetic monopoles. This phase has been denominated in the
literature the  He-McKellar-Wilkens(HMW) phase, and it is the
Maxwell dual of AC phase. A  simple practical experimental
configuration to test the HMW phase,
was proposed by Wei, Han and Wei\cite{wei} where the electric
field of a charged wire polarizes a neutral atom and an uniform
magnetic field is applied parallel to the wire. In this field
configuration, the atom acquires a HMW phase.  Dowling, Williams
and Franson\cite{fra} have proposed a unified description of
all three phenomena(AB, AC, HMW phase) discussing Maxwell
electromagnetic duality relation between the three quantum
phases. The analysis of the duality predicts a fourth
phenomenon, which is the dual of Aharonov-Bohm effect(dAB).
In a  recent article Anandan\cite{ana} has presented a unified
and fully relativistic treatment of the interaction between a particle with permanent electric and magnetic dipole moments and an electromagnetic field. This phase we designate Anandan topological phase. 

 In the early
eighties Berry discovered~\cite{ber} that a slowly
evolving(adiabatic) quantum
system retains information of its evolution when returnes to
its original
physical state. This information corresponds to what is
termed Berry's phase.
The appearance of this phase has been generalized to the case
of non-adiabatic~\cite{ahan} evolution of a quantum system. In any case the
phase only depends on the geometrical nature of the pathway along which the
system evolves.
This phenomenon has been investigated in several areas of
physics. There are
various experiments  which have been reported concerning the
appearance
of the adiabatic and non-adiabatic geometric phases,
including the
observations on photons\cite{tom}, neutrons\cite{bit} and
nuclear
spins\cite{tyc}. The manifestations of  Berry's quantum phase
in high-energy
electron diffraction in a deformed crystal, which contains a
screw
dislocation, has also  been observed \cite{bir}. 
The quantum evolution
of a single particle  is given by a time-dependent
Schr\"odinger equation $
  i\frac{d}{dt}|\psi(t)\rangle =\mathcal{H}
(R_{i}(t))|\psi(t)\rangle,
$
 where $H(R_{i}(t))$ is the system Hamiltonian
 which depends on  a  set of  externally controllable
parameters ${\bf R}(t)=
R_{i}$. The solution of the  Schr\"odinger equation in the
adiabatic
approximation is given by
$
 |\psi(t)\rangle =
\exp[-i\int_{0}^{t}{\cal E}(t)dt]\exp(i\gamma(C))|\psi({\bf
R}(t))\rangle ,
$
  where
$|\psi({\bf R}(t))\rangle$ are instantaneous eigenstates of
the Hamiltonian
with eigenvalues $ {\cal E}(t)$. The first factor of the phase is the usual
dynamical one.
The extra phase factor, $\exp(i\gamma(C))$, becomes physically
important when  the
parameters are changed along a path $C$ in the parameter
space,  over some time $T$, such as $R(T)=
R(0)$. The non-trivial phase factor is  Berry's  phase for
the  path $C$  which is given by
$
\label{berry}
 \gamma_{n}(C)=i\oint_{C} \langle\psi({\bf
R})|\nabla_{R}\psi({\bf R})\rangle d{\bf R},
$
 where  the term $\langle\psi({\bf
R})|\nabla_{{\bf R}}\psi({\bf R})\rangle$
 is denominated Berry connection.
Berry showed that the AB-effect is a special case of this
geometrical phase. In 1991, Mignani\cite{mig} and He and
McKellar\cite{mac1} also demonstrated by a different approach
that the AC-phase is a  particular case of Berry's quantum phase.

In this rapid communication we analyze  Anandan's phase for a quantum particle with  permanent magnetic and electric dipole moments in a radial configuration
of electric and magnetic fields and we study the appearance of a
geometrical quantum phase in their dynamics. We also demonstrate that Anandan's
topological phase is a special case of the Berry's geometrical
phase. 

We consider now the situation pointed by Anandan of
a neutral particle with permanent electric ($\mathbf{d}$) and
magnetic (${\bm\mu}$) dipole moments. The non-relativistic Lagrangian ($c=1$) that describes
this particle in the presence of an electromagnetic field is given by
\begin{eqnarray}
  \label{lag}
  \mathcal{L}=\frac{1}{2} m V^{2}  + \mathbf{d}\cdot(\mathbf{E} +
\mathbf{V}\times \mathbf{B}) + \bm{\mu}\cdot (\mathbf{B} -
\mathbf{V}\times \mathbf{E})
\end{eqnarray}
From (\ref{lag}) we obtain the canonical momentum,
\begin{equation}
  \label{can}
  \mathbf{P}=m\mathbf{V} - \bm{\mu}\times\mathbf{E} +
\mathbf{d}\times \mathbf{B},
\end{equation}
 and the equation of motion for the particle is given by
\begin{widetext}
\begin{eqnarray} 
  \label{em}
  m\ddot{\mathbf{r}}=\bm{\nabla}(\mathbf{\mu}\cdot\mathbf{B}
&+& \mathbf{d}\cdot
\mathbf{E})+\bm{\mu}\times\frac{\partial\mathbf{E}}{\partial
t} - \mathbf{d}\times \frac{\partial \mathbf{B}}{\partial t}
+\mathbf{V}\cdot \bm{\nabla}(\bm{\mu}\times \mathbf{E}) 
 - \mathbf{V}\cdot\bm{\nabla}(\mathbf{d}\times
\mathbf{B}) - \bm{\nabla}(\bm{\mu}\cdot \mathbf{V}\times
\mathbf{E})\\ \nonumber \\ \nonumber
&+&\bm{\nabla}(\mathbf{d}\cdot\mathbf{V}\times\mathbf{B})-\mathbf{\dot{d}}\times\mathbf{B}+\mathbf{\dot{\mu}}\times \mathbf{E}
\end{eqnarray}
\end{widetext}
We use  $\frac{d}{dt}=\partial_{t} +
\mathbf{v}\cdot\bm{\nabla}$ and the Maxwell equations
$\partial_{t}\mathbf{B}=-\bm{\nabla}\times\mathbf{E}$ and
$\partial_{t}\mathbf{E}=\bm{\nabla}\times\mathbf{B}$. We also
use the identities $\bm{\nabla}(\mathbf{d}\cdot
\mathbf{E})=(\mathbf{d}\cdot \bm{\nabla})\mathbf{E} +
\mathbf{d}\times (\bm{\nabla}\times \mathbf{E})$ and the  similar one
for the  magnetic case. After some mathematical manipulation, 
equation (\ref{em}) assumes the following form
\begin{eqnarray}
  \label{em1}
  m\ddot{\mathbf{r}}&=&(\mathbf{d}\cdot\bm{\nabla})\mathbf{E}
- \mathbf{v}\times\bm{\nabla}\times(\mathbf{d}\times
\mathbf{B}) + \dot{\mathbf{d}}\times\mathbf{B} \nonumber\\
&+& (\bm{\mu}\cdot\bm{\nabla})\mathbf{B} +
\mathbf{v}\times\bm{\nabla}\times(\bm{\mu}\times\mathbf{E})
-\dot{\bm{\mu}}\times\mathbf{E}.
\end{eqnarray}
For a complete description of the  dynamics of this particle we
need to find the  equation  that describes the dynamics of
dipoles. We assume that $\mathbf{d}$ and $\bm{\mu}$ refers to a permanent
electric and magnetic dipoles of a molecule or atom, respectively.
Hence, with the proviso of a subsequent justification, we set
$\dot{\mathbf{d}}=0$ and $\dot{\bm{\mu}}=0$, in Eq. (\ref{em1}),
and we consider that laboratory fields $\mathbf{E}$ and
$\mathbf{B}$ do not  vary in the dipole directions, and therefore $(\vec{d}\cdot\bm{\nabla})\mathbf{E}=0$ and
$(\bm{\mu}\cdot\bm{\nabla})\mathbf{B}=0$. In this paper we do
not consider the new contribution for the topological phase
found by Spavieri~\cite{spa}, since his contributions were null due to the field configurations adopted in this work. In this way,
the equation of motion is given by
\begin{eqnarray}
\label{f1}
 m\ddot{\mathbf{r}}=\mathbf{v}\times\bm{\nabla}\times[\bm{\mu}\times\mathbf{E}
-\mathbf{d}\times\mathbf{B}].
\end{eqnarray}
This force is reminiscent of the force experienced by an
electron moving in a magnetic field
$e^{-1}\bm{\nabla}\times[\bm{\mu}\times\mathbf{E}
-\mathbf{d}\times\mathbf{B}] $. 
The non-relativistic Hamiltonian for this problem can be obtained from the Lagrangian, and  is given by

\begin{eqnarray}
  \label{ham}
 \hat{\mathcal{H}}=\frac{1}{2m}(\hat{\mathbf{p}} - {\bm \mu}\times{\bf
E} + {\bf d}\times {\bf B})^{2} - {\bm\mu}\cdot{\bf B} - {\bf d}\cdot{\bf E}.
\end{eqnarray}
Let us analyze the quantum dynamics of a particle governed by the  Hamiltonian(\ref{ham}). In the region in which the particle is immersed, we consider that the fields are cylindrically radial. We adopt the following generic fields

\begin{eqnarray}
\label{gen}
\mathbf{B}=\frac{\lambda_{m}}{\rho} \hat{e_{\rho}} \quad \mathbf{E}=\frac{\lambda_{e}}{\rho}\hat{e_{\rho}},
\end{eqnarray}
\noindent
where $\lambda_{m}$ is a constant that depends on the  array of
magnetic  dipoles or magnetic charges that generate a magnetic field and $\lambda_{e}$ is an electric charge density
that creates the electric field.
With this field configurations the expression  for the  force
experienced by the particle, (Eq.(\ref{f1})) is null. If we
analyze the expression for the torque experienced by the
dipoles in this field arrangement we can see that it
vanishes. In this way the conditions for the  existence of
topological phase can be assured\cite{ana}. The potential term
in the Hamiltonian (\ref{ham}) gives null contribution to  the
dynamics of this dipole in the field configurations
(\ref{gen}) if the dipoles are prepared to be  aligned with the
$z$-direction. We assume that the particle is moving in the $x-y$  plane , in an
external electric and magnetic field. We also suppose that the fields
generated by the sources are radially distributed in the space.
We adopt a treatment analogous to the  Aharonov-Bohm effect
as used by Berry\cite{ber}. In this framework, Berry's phase can be calculated by use of the following arguments: let us confine the quantum  system to a perfectly reflecting  box such that the wave packet is nonzero only in
the interior of the box.  The vector that localizes  the box in relation to
the source  of electromagnetic  field
is called  ${\bf R}$. This vector is oriented from the origin
of the
coordinate system (localized on the source) to the center of
the box.
 In the
absence of fields,  the wave function for the particle in the
box  is given by
$\psi_{0}({\bf r}-{\bf R})$ where ${\bf r}$ represents  the
coordinates of the particle centered at {\bf R}. 
We use the Dirac phase factor method to describe the quantum 
dynamics of the particle. In this approach the quantum state of the particle in the presence of these electromagnetic fields can be written as a  function of
the  quantum state in the absence of the electromagnetic field
times a multiplicative exponential term that depends on the
fields. The quantum dynamics is  governed by the  time-dependent
Schr\"odinger equation, and the quantum state can be written as
%\begin{widetext}
\begin{eqnarray}
  \label{dir}
   |\Psi(t)\rangle=
\exp\left[-i\int_{0}^{t}{\cal E}(t)dt\right]\times\nonumber\\
\exp\left[-i\int_{R}^{r'}(\bf{d}\times\bf{B}
-\bm{\mu}\times\bf{E})dr'\right]|\Psi_{0}({\bf R}(t))\rangle,
\end{eqnarray}
%\end{widetext}
where $|\psi_{0}(\rho,\phi)\rangle$ is the solution of the
Schr\"odinger equation in absence of fields. The first term
in (\ref{dir}) is the dynamical phase and the second
one is the Berry's geometrical phase. If we
transport the box around the sources of the electromagnetic
fields,  the  wave function of the particle acquires a
Berry's quantum  phase. Then,  in order to  compute
this, we need to calculate the Berry's connections
\begin{widetext}
\begin{eqnarray}
  \label{conec}
A_{ij}&=&  \langle\Psi({\bf r}-{\bf
R})|\nabla_{\bf{R}}(\Psi({\bf r}-{\bf R})\rangle
=
\int d^{3}{\bf r}\Psi^{*}(\bf{r}-\bf{R})[-i(\bf{d}\times\bf{B}
-\bm{\mu}\times\bf{E})\Psi(\bf{r}-\bf{R}) +
\nabla_{\bf{R}}\Psi(\bf{r}-\bf{R})].
\end{eqnarray}
\end{widetext}
Which  results in
\begin{eqnarray}
  \label{int}
 A_{ij}=i(\bf{d}\times\bf{B} -\bm{\mu}\times\bf{E}).
\end{eqnarray}
 
Thus, the Berry's phase is given by
\begin{eqnarray}
   \label{fas}
   \gamma_{c}(C)= \oint (\bf{d}\times\bf{B}
-\bm{\mu}\times\bf{E})d\bf{R}.
\end{eqnarray}

The effect is observed as an interference between the
particle in the
transported box and one in a box which was not transported
around the
circuit. This result demonstrates  the existence of Berry's
quantum phase in the
dynamics of a neutral particle that contains a permanent
electric and magnetic moments of dipole. 
 Our purpose at this  moment is to introduce some setups in which we obtain certain field and dipole moment configurations where this phase can be observed.  

The first set-up that we analyze is the radial field configuration presented in (\ref{gen}). The arrangement of field configurations with the magnetic field radially cylindrical  is more difficult to achieve experimentally, due to the fact that, a priori, we need a linear distribution of magnetic charges. We can observe in the literature that this kind of arrangement would be possible. Some authors have claimed that this configuration can be obtained experimentally as in the arrangements presented  in the articles\cite{mag,mag1,mag2}. 
In this experimental set, the radial magnetic field is produced by two concentric cylindrical magnets, and in this way a  radial magnetic field can be produced between the magnets.  More realistic configuration  that can generate a radial magnetic field is when  we consider a cylindrical solenoid carrying a surface current density that produces, in the  external region of a solenoid, a field configuration with the  same characteristics of (\ref{gen})~\cite{chys}. Despite all difficulties to do a real experiment, we suppose that it is possible. If we cogitate in this region the existence of  a radial electric field generated by an internal cylindrical electrode, we verify that Berry's quantum phase (\ref{fas}) has the following expression
\begin{eqnarray}
  \label{confi}
   \gamma_{c}(C)= 2\pi\mu \lambda_{e} -2\pi d \lambda_{m}.
\end{eqnarray}
Other device of radial magnetic and electric fields is
obtained if we use the same arguments presented by Tkachuk~\cite{tka}.
In that case, he has considered that the radial field configurations are generated by a charged ferromagnetic wire in the  $z$-direction. We
use this arrangement with a simple difference, namely we adopt here that the ferromagnetic wire is electrically charged. The magnetization is parallel to the wire and its magnitude is $M(z)=-\lambda_{m}z$, where
$\lambda_{m}$ can be treated  as a linear magnetic charge density.  Additionally this wire has a linear electric charge density that generates a radial electric field. When the wire
is sufficiently long, the magnetic and electric fields around
the central part of the  wire are
\begin{equation}
  \label{cam}
  {\bf B} = \frac{2\lambda_{m}}{\rho}\hat{e_{\rho}} \quad
{\bf E}=\frac{\lambda}{\rho} \hat{e_{\rho}}.
\end{equation}
Berry's quantum phase (\ref{fas}) for the modified Tkachuk's configuration is given by the following phase factor 
\begin{eqnarray}
  \label{confi1}
   \gamma_{c}(C)= 2\pi\mu \lambda_{e} -4\pi d \lambda_{m}.
\end{eqnarray}
For the field configurations analyzed   in the present letter we
observe that: for the case  $\lambda_{e}=0$,  obtained from Eq.(\ref{fas}),
the HMW phase is a special case of Berry's
phase. In the case  $\lambda_{m}=0$, the results of 
Mignani\cite{mig} and He and McKellar\cite{mac} for the
Aharonov-Casher effects are obtained. In the case 
$\lambda_{e}\neq 0$ and $\lambda_{m}\neq 0$, we obtain  
Anandan's generalized topological phase for a quantum
particle (atom or molecule), with a permanent magnetic
and electric dipole moments.  This is a special case of the
geometric Berry's quantum phase.
\begin{figure}
\begin{center}
\resizebox{8cm}{3cm}{\includegraphics{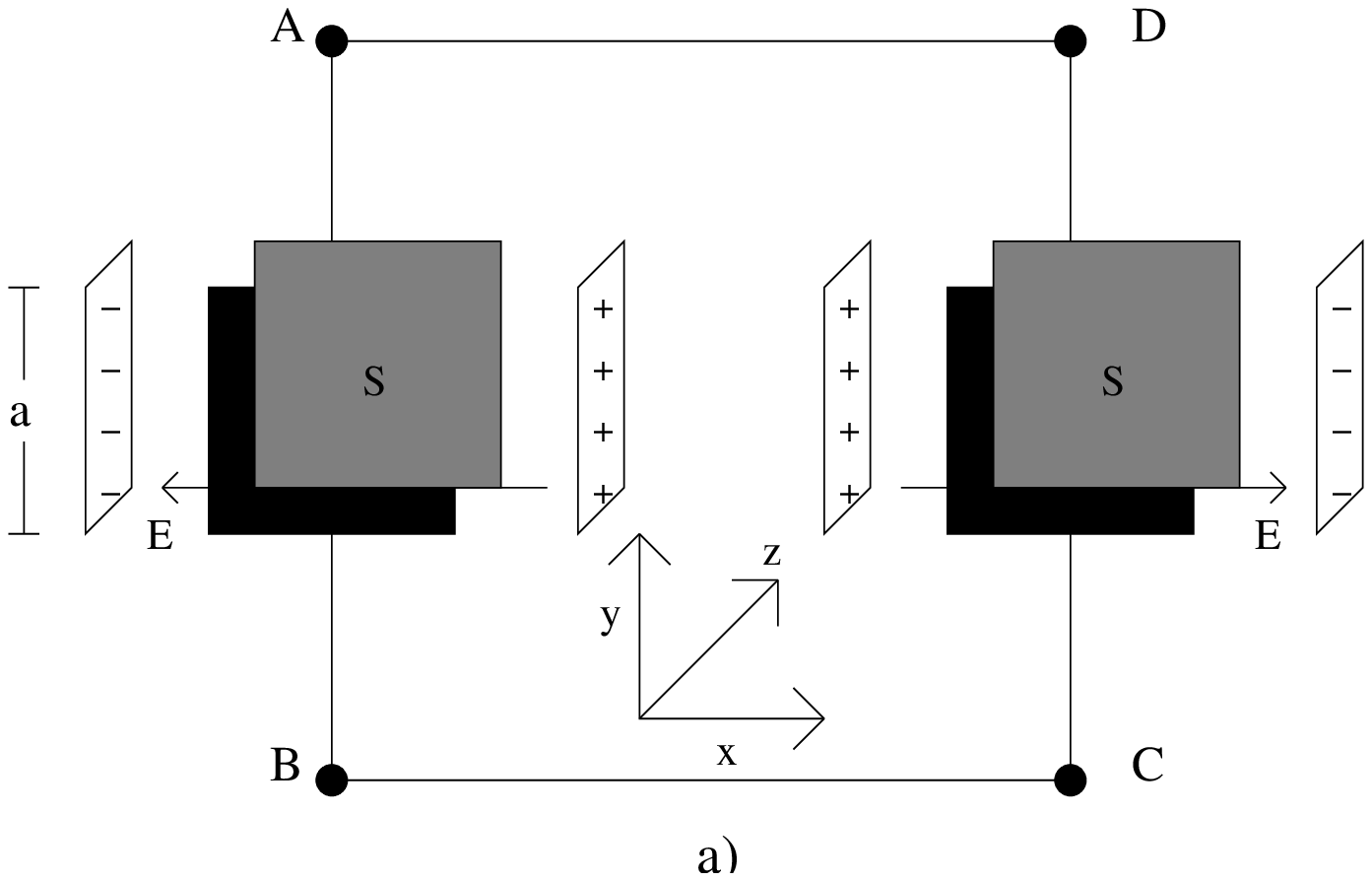} \hspace{2cm}\vspace{0.5cm}\includegraphics{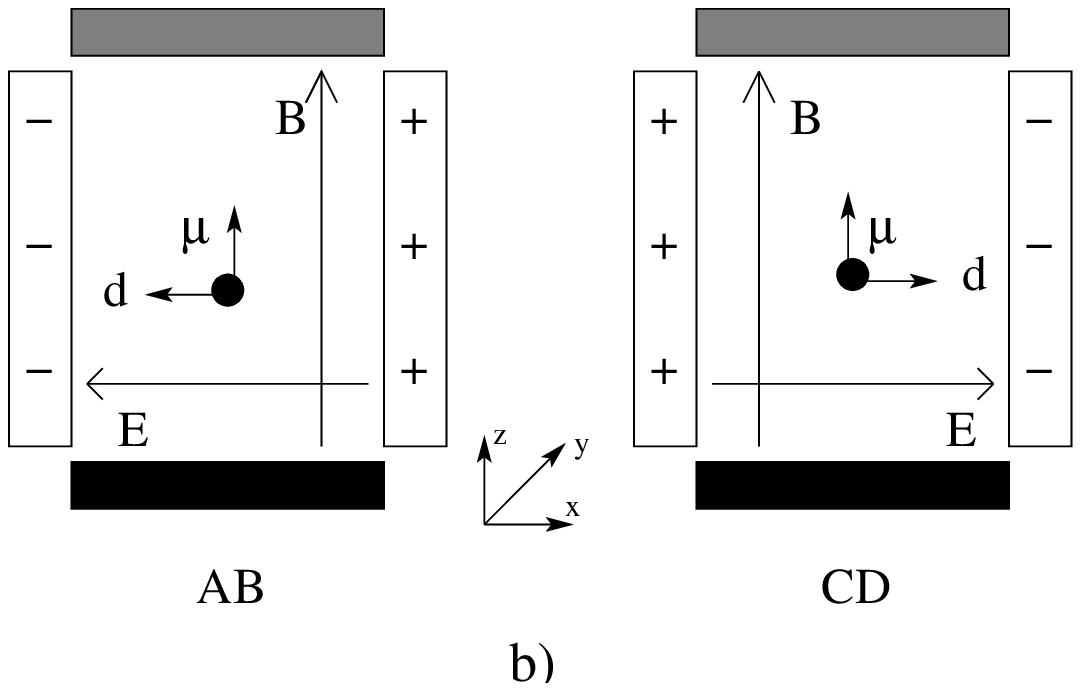}}   
\caption{a)Schematic diagram for obtaining Anandan's phase. We assume that beam of  particles are prepared with magnetic moment aligned with the{\bf z}-direction and are introduced in a region with velocity in {\bf y}-direction,  where a constant magnetic field is also present in the {\bf z}-direction. We also have a electric field in {\bf x}-direction. b)Cross section where we can see the dipole alignments.}
\label{fig1}
\end{center}
\end{figure}

Now, the analysis of the Berry's phase in this system is developed in the  following way: we are interested in finding configurations of {\bf d}, {$\bm\mu$}, {\bf B} and {\bf E} that make the classical torque and force to vanish. Recently Lee\cite{lee} has studied the quantum phase in dynamics of the electric or  magnetic dipoles and he observes that it is possible to write a vector potential for magnetic or electric dipole using the point of view of the Aharonov Bohm effect, noting that there is  a dual effect of the Casella\cite{case} effect for dipole. The setup introduced by Lee, has guided our third configuration for a very simple apparatus for observation of the  quantum phase in the present case. We can find that the electric dipole, in this apparatus, is orthogonal to {\bf B} and  {\bf v} at the same time. In the  same way if  the magnetic dipole is perpendicular to {\bf E} and {\bf v}, it has the  property of vanishing both, torque and force. The quantum phase acquired by both dipoles is only independent of the trajectory in opposition to the Aharonov-Bohm phase that is independent of the trajectory and is also independent of the solenoid configuration. The trajectory of the dipoles is restricted and the velocity should be perpendicular to  both moments because of the relativistic effects in the fields. We consider a schematic experiment showed in  figure(\ref{fig1}). We assume that a beam of atoms or molecules, with permanent magnetic dipole moments and  that can  induces electric dipole moments in it, is prepared with the magnetic dipole moment  aligned with {\bf z}-direction and the electric dipole moment  aligned with {\bf x}-direction. This  beam is introduced in the interferometer device with a set of  two dipole magnets with the magnetic field in {\bf z}-direction and two generators of uniform electric field in {\bf x}-direction but in opposite direction as showed in  figure (\ref{fig1}). The resultant phase acquired by atoms (or molecules) in this interferometer paths is given by
 \begin{eqnarray}
   \label{casela}
   \gamma(c)= 2dB_{0}a + 2\mu E_{0}a
 \end{eqnarray}
where $B_{0}$ and $E_{0}$ are the magnitude of the fields in both path around the apparatus and $a$ is the length of both, magnets and electric plates. For $E_{0}=0$ we have the dual Casella effect predicted by Lee; for $B_{0}=0$ we have the Casella effect. This is a generalized Casella effect and also is a  case of the Anandan phase.

We have shown that the Anandan quantum phase is a special case of  Berry's phase. Our purpose is to configure a process similar to an interferometer to try to observe this quantum phase. If we  can be able to do it, we hope that neither a classical force nor a classical torque will act in the test particle(which can be an atom or molecule). The interference framework should be explained only by a quantum potential effect in the sense of the Aharanov-Bohm effect. We have demonstrated three configurations where the geometric phase can be detected. In the third configuration we have obtained a generalized Casella effect that is a case of a class of phases treated in this work. 

\acknowledgments
This work was partially supported by CNPq and by Funda\c{c}\~ao de Amparo \`a Pesquisa do Estado da Bahia(FAPESB).


\begin{thebibliography}{99}

\bibitem{aha} Y. Aharonov  and D. Bohm, Phys. Rev. {\bf 115}, 485 (1959).

\bibitem{prl:cham}R. G. Chambers, Phys. Rev. Lett. {\bf 5},3 (1960). 

\bibitem{pes}M.  Peshkin  and A. Tonomura, {\it The Aharonov-Bohm Effect} (Springer-Verlag, Berlin, 1989).

\bibitem{cas} Y Aharonov and A. Casher, Phys. Rev. Lett. {\bf 53}, 319, (1984).

\bibitem{cim} A. Cimmino et al., Phys. Rev. Lett. {\bf 63}, 380 (1989).

\bibitem{san} K. Sangster et al., Phys. Rev. Lett. {\bf 71},3641 (1993).

\bibitem{san1} K. Sangster et al., Phys. Rev. A {\bf 51}, 1776 (1995).
\bibitem{mac} X. -G. He and B. H. J. McKellar, Phys. Rev. A, {\bf 47}, 3424 (1993).

\bibitem{wil}M. Wilkens, Phys. Rev. Lett. {\bf 72}, 5 (1994).

\bibitem{wei}H. Wei, R. Han and X. Wei, Phys. Rev. Lett. {\bf 75}, 2071 (1995).

\bibitem{fra}J. P. Dowling, C. P. Williams and J. D. Franson, Phys. Rev. Lett. {\bf 83}, 2486(1999).

\bibitem{ana}J. Anandan, Phys. Rev. Lett. {\bf 85}, 1354 (2000).

\bibitem{ber} M. V. Berry, Proc. R. Soc. London, {\bf A 392}, 457 (1984).

\bibitem{ahan}Y. Aharonov and J. Anandan,  Phys. Rev. Lett. {\bf 58}, 1593 (1987).

\bibitem{tom} A. Tomita and  R. Y. Chiao, Phys. Rev. Lett. {\bf 57}, 937 (1986) ; R. Simon, H. J. Kimble  and E. C. G. Sudarshan, Phys. Rev. Lett. {\bf 61}, 19 (1988)

\bibitem{bit} T. Bitter  and D.  Dubbers, Phys. Rev. Lett. {\bf 59}, 251 (1987) .

\bibitem{tyc} R. Tycko, Phys. Rev. Lett. {\bf 58}, 2281 (1987).

\bibitem{bir} D. M. Bird and A. R. Preston, Phys. Rev. Lett. {\bf 61}, 2863 (1988).

\bibitem{mig}R.  Mignani, J. Phys. A: Math. Gen. {\bf 24}, L421 (1991).

\bibitem{mac1}X. -G. He and B. H. J. McKellar, Phys. Lett. B {\bf 264} 129 (1991)

\bibitem{spa}G. Spavieri, Phys. Rev. Lett. {\bf 82}, 3932 (1999).

\bibitem{mag}W. H. Heiser and J. A. Shercliff, J. Fluid Mech. {\bf 22}, 701 (1985).

\bibitem{mag1}S. Y. Molokov and J. E. Allen, J. Phys. D: Appl. Phys. {\bf 25}, 393 (1992).

\bibitem{mag2}S. Y. Molokov and J. E. Allen, J. Phys. D: Appl. Phys. {\bf 25}, 933 (1992).

\bibitem{chys} C. Chyssomalakos, A. Franco and A. Reyes-Coronado, {\it ``Spin $\frac{1}{2}$ particle on a Cylinder with Radial Magnetic Field''}, quant-ph/0212054 (2002).

\bibitem{tka} V. M. Tkachuk, Phys. Rev. A {\bf 62}, 052112 (2000). 

\bibitem{lee} T. -Y. Lee, Phys. Rev. A {\bf 64} 032107 (2001).

\bibitem{case} R. C. Casella, Phys. Rev. Lett. {\bf 65}, 221 (1990).
\end{thebibliography}
\end{document}